\documentstyle[preprint,epsfig,prb,tighten,aps]{revtex}

\newcommand{\eqn}[1]{Eq.\ (\ref{#1})}

\begin{document}
\draft

\preprint{G\"oteborg ITP 98-10}
\title{Landau Ginzburg theory of the d-wave Josephson junction}
\author{ Stellan \"Ostlund}
\address{
Institute of Theoretical Physics\\
Chalmers University of Technology\\
S-41296 G\"{o}teborg, Sweden}
\date{\today}
\maketitle
\begin{abstract}
This letter discusses the Landau Ginzburg theory of a Josephson 
junction composed of on one side a pure d-wave superconductor  oriented
with the $ (110)$ axis normal to the junction and on the other side
either s-wave or d-wave oriented  with $(100) $ normal to the junction.
We use simple symmetry arguments  to show that the Josephson current as
a function of the phase must have the form 
$ j( \phi ) = j_1    \sin( \phi ) + j_2  \sin( 2 \phi ) $.  
In principle $ j_1 $ vanishes for a perfect junction of this type,
but anisotropy effects,  either due to a-b
axis asymmetry or junction imperfections  can easily cause $ j_1 /
j_2 $ to be quite  large even in a high quality junction. If $ j_1 / j_2  $ is
sufficiently small and $ j_2 $ is negative, local time reversal symmetry breaking will 
appear.
Arbitrary values of the flux would then be pinned to corners between
such junctions and occasionally on junction faces, which is consistent with 
experiments on grain boundary junctions.

\end{abstract} 

\pacs{PACS 74.50.+r, 74.76.Bz, 74.80.Fp, 74.20.De}

A Josephson junction that is constructed with d-wave superconductor
oriented with the (110) axis  normal to the junction  and  either  an
s-wave superconductor  or a  d-wave superconductor oriented with the
(100) axis normal to the junction has a reflection symmetry in the plane
of the junction which suppresses the conventional Josephson coupling.
Calculation involving all orders of the tunneling process shows that
this leaves  an anomalous part that is doubly periodic in the Josephson
phase.  \cite{Annett,Tanaka94,Furusaki} This effect has been further explored 
by a number of authors through the Bogoliubov-de Gennes equations and 
interpreted in terms of Andreev levels.  \cite{Zago,Fogelstroem:97a,cth}

A number of recent papers have independently explored the possibilities 
of exotic local order such as 
$ ( s + i d ) $ or $ d_{x^2 - y^2 } + i d_{xy} $ appearing
near a bicrystal junction.  These latter scenarios have
suggested that coupling to these order parameters can be important
in the vicinity of the junction and can lead to spontaneously broken 
time reversal near the junction.  Different couplings 
and mechanisms have been proposed to explain the effect. 
\cite{Sigrist:95a,Huck97,Bailey97}

This paper shows that symmetry alone forces
all these features to be present in the Landau Ginzburg theory of
a pure $ d_{x^2 - y^2} $ superconductor without assuming the existence 
of any other local order parameters or any specific mechanism.

The geometry of the junction we will consider is given in Fig.[1].  Our
detailed discussion will assume there are identical  d-wave superconductors on
both sides, but the arguments apply equally well to an asymmetric
junction with s-wave on the left side and a d-wave on the right. (Fig.[2]) 
We will consider system with a uniform junction along the y-axis so the
only spatial dependence is along x.

The free energy of the  superconductor is given by  a bulk term 
$ {\cal F } =  {\cal F }_{bulk}  + {\cal F }_{jct} $ where

\begin{equation}
{\cal F }_{bulk} = \int_{-\infty}^{\infty} (  - | { \psi(x) } |^2 +  
	{\scriptstyle{ \frac{1}{2}}} \;  | { \psi(x) } |^4 +  | { { d \psi / dx } |}^2 
\end{equation} 
with the important terms in the junction energy given by 
$ {\cal F }_{jct} =  \int \delta(x) f_{jct} dx $ where $ f_{jct} = $ 
\begin{equation}
\alpha \left( | {\psi_+} |^2 +   | {\psi_-} |^2 \right) -
     \beta ( \psi_+^* \psi_- + CC ) -
      {\scriptstyle{ \frac{1}{2}}} \gamma \left( ( \psi_- \psi_+^*)^2 + CC  \;  \right)
\end{equation}

We have normalized the length in units of coherence length, and the order 
parameter to its value at $ \pm \infty $. We use the notation
$ \psi_\pm  =   { lim_{x \rightarrow 0_\pm}} \psi( x ) $ so that   $ \psi_\pm $ 
indicates the order parameter on either side of the
junction.   Other fourth order terms at the junction such as $ | { \psi_- \psi_+ } |^2 $ can be 
neglected assuming ``$ \alpha  $'' is  large compared to $ | { \psi_\pm} |^2 $.

The Josephson equations are then
\begin{equation}
\begin{array}{llll}
0  = 
& - &  { {d^2}\over{d^2x}} \psi(x) -  \psi(x) +  | {\psi(x)} |^2 \psi(x)  \\
0  = 
& -  &\left( { d\over{dx}} \psi ( x ) \right)_{x \rightarrow 0_-} + 
	\alpha  \psi_- - \beta \psi_+ - \gamma \psi^*_- ( \psi_+)^2   \\
0  = 
&   & \left( { d\over{dx}} \psi ( x ) \right)_{x \rightarrow 0_+} + 
	\alpha  \psi_+ - \beta \psi_- - \gamma \psi^*_+ ( \psi_-)^2   
\end{array}
\end{equation}
where the second and third term include a  derivative from integration by parts in addition to terms explicitly derived
from the junction  free energy.

For most weakly coupled junctions, the ``$ \gamma $'' term can safely be ignored, since $ | {\psi_\pm} |^2 $
in the junction is small and the $ \beta $ term will totally dominate the Josephson
coupling. However, when the junction has the symmetry of Fig. [1] or Fig [2],  the
junction becomes symmetric under reflections $ y \rightarrow -y $. For this symmetry operation
$ ( \psi_- , \psi_+)  \rightarrow ( \psi_- , - \psi_+ ) $ so that in this
case $ \beta $ must vanish and the fourth order term will determine the residual 
coupling. \cite{Annett}

For a symmetric junction, we can take $ \psi_+ = e^{ i \phi } \psi_- $; multiplying
the second term by $ \psi^* _- $ and taking the imaginary part, we find that the Josephson current-phase
relation is given by
\begin{equation} \label{eqn:phasecur}
j ( \phi )  =  j_1 \sin( \phi ) + j_2 \sin( 2 \phi )  \\ 
\end{equation}
where
\begin{equation} \label{eqn:ratioj1j2}
j_1 / j_2  = \beta / (  \gamma | {\psi_-} |^2 )
\end{equation}
Demanding that $ j ( \phi ) = 0 $ implies that either $ \phi = 0,\pi $ or
$ \cos( \phi_c ) = - \beta / ( 2 \gamma | { \psi_+ } |^2 ) $. If the ordinary Josephson coupling
term $ \beta $ is sufficiently small, this equation will have a symmetric pair of nontrivial solutions.
In the limit of weak Josephson current, we can approximate the boundary condition governing 
the magnitude of $ \psi $ by $ - { d\over{dx}}  | { \psi_- } | +  \alpha  | {\psi_-} | = 0 $.

Comparing the junction free energies of the three solutions we find that  $  \delta {\cal F }_{jct} $,
the portion of the junction energy that varies with $ \phi $ are
\begin{equation} \label{eqn:phic}
\begin{array}{l}
 \delta {\cal F }_{jct} ( \phi = 0  )  = | {\psi_-} |^2 (  - 2 \beta - \gamma | {\psi_-} |^2 ) \\
 \delta {\cal F }_{jct} ( \phi = \pi )  = | {\psi_-} |^2  (   2 \beta - \gamma | {\psi_-} |^2 ) \\
 \delta {\cal F }_{jct} ( \phi = \phi_c )  =   \beta^2 /( 2 \gamma ) + \gamma | {\psi_-} |^4 
\end{array}
\end{equation}

We see that if $ \gamma \; < \; 0 $, the anomalous solutions  
$ \phi_c $ will dominate when $ \beta $ is sufficiently small and the
junction will favor a zero-current state with broken time reversal
symmetry. In the limit $ \beta =  0 $ the current is $ \pi $-periodic
rather than $ 2 \pi $-periodic as would be expected in a conventional
Josephson junction.\cite{Annett,Tanaka94,Furusaki,Zago}
We note that these conclusions hold also for a junction where
one side of the junction is a d-wave superconductor with the junction
along the (110) axis and the other side of the junction is pure s-wave.

For a Josephson junction without the special inversion symmetry, the
``$\beta$'' term does not vanish by symmetry. However, when the
relative crystal orientations  are close to $ \pi/4 $ but microscopic
arguments indicate that it may still be  small.\cite{Annett,reviews}
Whether or not this is actually the case can in principle be measured
in a phase-current experiment.

In the experiments of Kirtley et al,\cite{Kirtley95} triangular and
hexagonal inclusions of crystal whose (100) is axis rotated by $ \pi / 4  $
relative to the surrounding crystals had apparently arbitrary values of
flux condensed at the corners.  Our scenario suggests that for each
such junction face there is a phase shift  across each interface given
by   different values of $ \phi_c $, which can take arbitrary values
across the interface if $ \gamma \; < \; 0 $. The value of the
flux condensed at each corner is simply the difference of the values of
$ \phi_c $ across the two faces that meet at the corner. There is also
the possibility of flux being pinned at an arbitrary location on the
interface boundary corresponding to a defect where the Josephson phase
changes between the two degenerate solutions. The flux trapped on such
a face can be any of the values  $ 2 n \pi \pm   2 \phi_c $.

Whether or not this scenario can explain Kirtleys experiments can be
further tested experimentally through phase-current measurements across
individual junctions with the same junction geometry that
occur in individual faces in  the Kirtley experiments.  In order to be
consistent with the above scenario, a triangular inclusion that shows
nonintegral flux condensation at two or more corners must have at least
one face that independently shows a Josephson current with sufficiently
large deviations from $ \sin( \phi ) $ behavior to have zero current at
two values of phase in addition to the solutions that are multiples of
$ \pi $ that occur since the Landau Ginburg equations are time reversal
symmetric.

A phase-current experiment such as the one recently made by  Il'ichev
\cite{Ilichev98} measures precisely \eqn{eqn:phasecur}.
\eqn{eqn:ratioj1j2} suggests, however, that  the most dramatic
deviations from pure sinusoidal behavior could be quite difficult to
reproduce from sample to sample, because the ratio  $ j_1 / j_2 $
depends on properties such as the barrier width and deviations from
inversion symmetry in the junction plane which could be hard to
control.  The {\em tendency } of the system to generate broken local
time reversal symmetry however, should be measurable through the sign
of $ j_2 $ which must be negative for such symmetry breaking to occur.

\acknowledgments The author particularly thanks
Zdravko Ivanov for constructive comments. This research was  sponsored 
by NFR, the Swedish National Research Council.

\begin{figure}
\begin{center}
\mbox{\epsfig{file=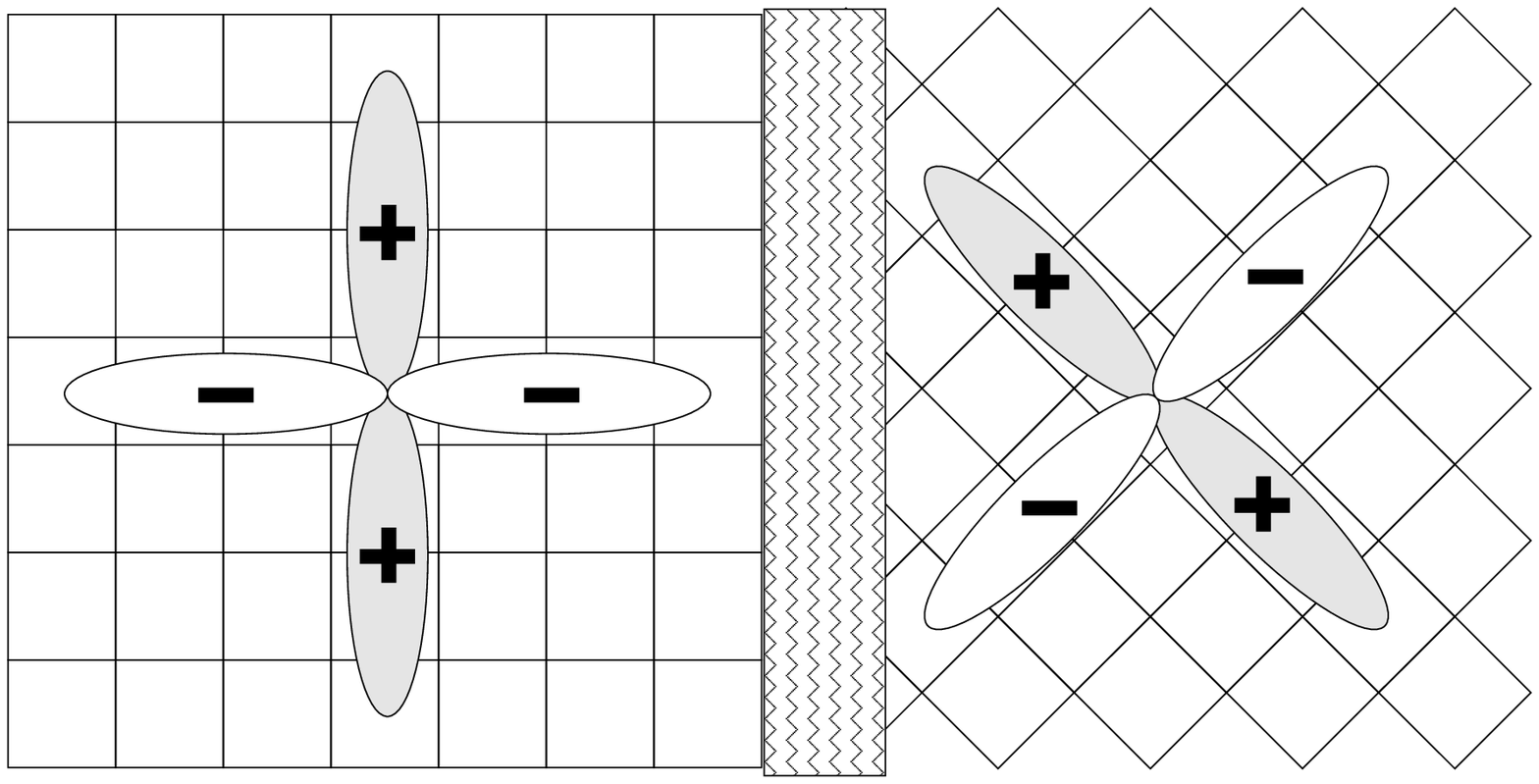,width=5.35in,angle=0}}
\end{center}
\caption{ The geometry of a symmetric $ ( 0, \pi/4 ) $  
Josephson junction. The system is symmetric under reflections 
through the x axis while the order parameter on the right changes sign.}
\end{figure}

\begin{figure}
\begin{center}
\mbox{\epsfig{file=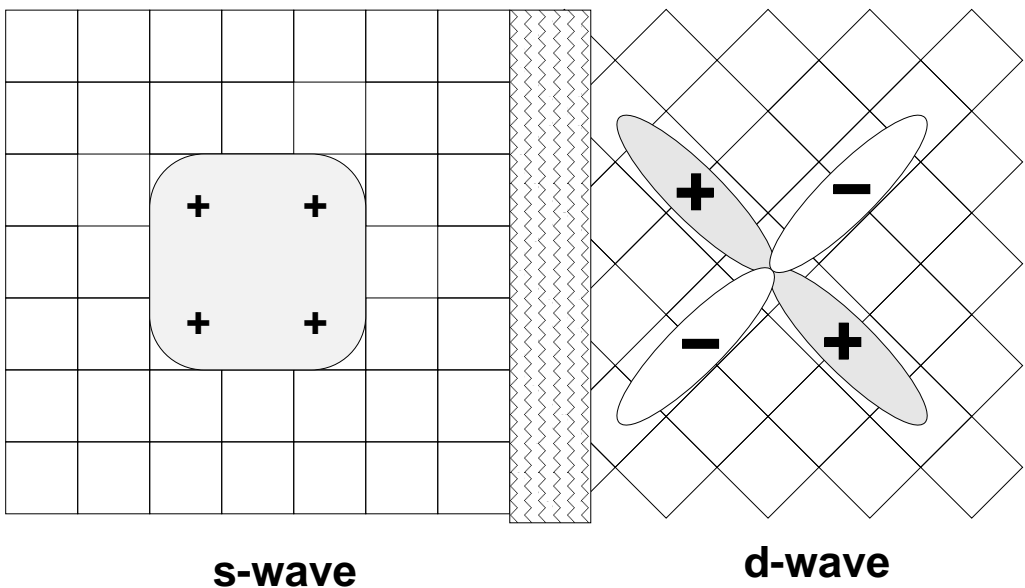,width=5.35in,angle=0}}
\end{center}
\caption{ An s-d Josephson junction with reflection symmetry through the x-axis.}
\end{figure}


\begin{references}

\bibitem{Kirtley95} Kirtley\ J.  et al: {\it Phys. Rev. B.} 51, 12057 (1995).
\bibitem{Annett} J.F.~Annett, {\it Adv. Phys. }{\bf 39} 83, (1990).
\bibitem{Tanaka94} Y. Tanaka,{\it Phys. Rev. Lett.}{\bf 24 },3871 (1994).
\bibitem{Furusaki} A. Furusaki and M. Tsukada, {\it Solid State Commun. }{\bf 78} 299 (1991).
\bibitem{Zago}A. M. Zagoskin, {\it J. Phys.: Condens. Matter } {\bf 9}, L419 (1997).
\bibitem{Fogelstroem:97a} M. Fogelstr\"{o}m, D. Rainer, and J.A. Sauls, {\it Phys. Rev. Lett. } {\bf 79}, 281 (1997); M.~Fogelstr\"{o}m, S. Yip, and J.Kurkj\"arvi, {\it Physica C } {\bf 294 }, 289 (1998).
\bibitem{cth} M. Hurd, {\it Phys. Rev. B. } {\bf 55 }, R11 993 (1997); T. L{\"o}fwander et al, {\it Phys. Rev. B. }{57 }, R3225 (1998);
\bibitem{Sigrist:95a} M.~Sigrist, D.~B.~Bailey, and R.~B.~Laughlin, {\it Phys.~Rev.~Lett.} {\bf 74}, 3249 (1995).
\bibitem{Huck97} A. ~Huck, A. ~van Otterloo and M. ~Sigrist, {\it  Phys.~Rev.~B } {\bf  56}, 14163 (1997).
\bibitem{Bailey97} D.~B.~Bailey, M.~Sigrist and R.~B.~Laughlin, {\it Phys. Rev. B} {\bf 55}, 15239 (1997).
\bibitem{reviews} M. Sigrist, T.M. Rice, {\it Rev. Mod. Phys.},{\bf 67 }, 503, (1995); D.J. Van Harlingen, {\it Rev. Mod. Phys.},{\bf 67 }, 515 (1995).
\bibitem{Ilichev98} E. Il'ichev, et al, {\it Phys. Rev. Lett. } to be published (1998).



%
%
%
%
%
%


\end{references}
\end{document}